\documentclass[reprint, preprintnumbers, superscriptaddress, nofootinbib, amsmath,amssymb, aps, prd, floatfix]{revtex4-2}
    
    \usepackage{graphicx}
    \usepackage[caption=false]{subfig}
    \usepackage[usenames,dvipsnames]{xcolor} 
    \usepackage{times, mathptmx} 
    \usepackage[utf8]{inputenc}
    \usepackage{color}
    \usepackage{hyperref}
    \usepackage[normalem]{ulem} 
    \usepackage{physics}
    \usepackage{comment}
    \usepackage{bm}
\hypersetup{
  colorlinks   = true, 
  urlcolor     = blue, 
  linkcolor    = blue, 
  citecolor   = blue 
}
    \allowdisplaybreaks[1]

\usepackage{float}

\newcommand{\mbh}{M_\mathrm{BH}}
\newcommand{\rmd}{\mathrm{d}}

\newcommand{\qmul}{Geometry, Analysis and Gravitation, School of Mathematical Sciences, Queen Mary University of London,
Mile End Road, London E1 4NS, United Kingdom}
\newcommand{\oxford}{Astrophysics, University of Oxford, Denys Wilkinson Building, Keble Road, Oxford OX1 3RH, United Kingdom}
\newcommand{\illinois}{Departments of Physics and Astronomy, University of Illinois at Urbana-Champaign, Urbana, IL 61801, USA}

\begin{document}

\title{The effect of wave dark matter on equal mass black hole mergers}

\author{Josu C. Aurrekoetxea}
\email{josu.aurrekoetxea@physics.ox.ac.uk}
\affiliation{\oxford}
\author{Katy Clough}
\affiliation{\qmul}
\author{Jamie Bamber}
\affiliation{\illinois}
\author{Pedro G. Ferreira}
\affiliation{\oxford}

\begin{abstract}
For dark matter to be detectable with gravitational waves from binary black holes, it must reach higher than average densities in their vicinity. In the case of light (wave-like) dark matter, the density of dark matter between the binary can be significantly enhanced by accretion from the surrounding environment. Here we show that the resulting dephasing effect on the last ten orbits of an equal mass binary is maximized when the Compton wavelength of the scalar particle is comparable to the orbital separation, $2\pi/\mu\sim d$. The phenomenology of the effect is different from the channels that are usually discussed, where dynamical friction (along the orbital path) and radiation of energy and angular momentum drive the dephasing, and is rather dominated by the radial force (the spacetime curvature in the radial direction) towards the overdensity between the black holes. Whilst our numerical studies limit us to scales of the same order, this effect may persist at larger separations and/or particle masses, playing a significant role in the merger history of binaries.
\end{abstract}

\maketitle


\textbf{\textit{Introduction.}}— Gravitational-wave observations provide a unique window that can be used not only to infer the astrophysical properties of black holes (BHs), but also to gather information about the environments they live in. The presence of matter around BHs during a binary merger event results in modifications to the trajectories, which in turn changes the gravitational-wave signal in a characteristic way 
\cite{Barack:2018yly,Barausse:2020rsu,LISAConsortiumWaveformWorkingGroup:2023arg,CanevaSantoro:2023aol,Cardoso:2019rou,Barausse:2014tra,Yunes:2011ws,Kocsis:2011dr,Macedo:2013qea,Cardoso:2020lxx,Bertone:2018krk,AlvesBatista:2021gzc,Zwick:2021dlg,Cardoso:2019upw,Maselli:2021men,Amaro-Seoane:2012lgq,Cardoso:2022whc,Cole:2022yzw,Krolak:1987ofj}.
Environments may arise from standard matter, such as accretion discs, or from dark matter (DM). In this work we focus on the latter case.

\begin{figure}[t]
    \centering
    \includegraphics[width=\linewidth]{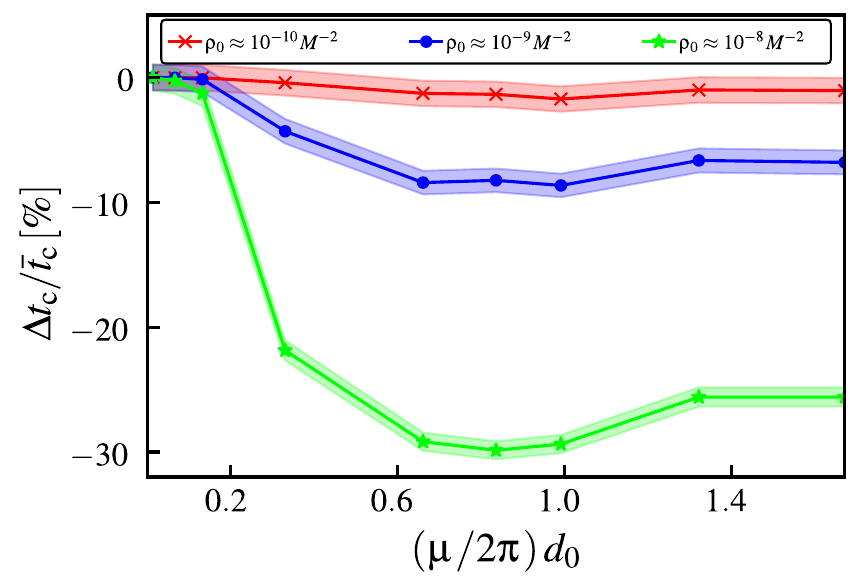}
    \caption{Dephasing in the coalescence time $\Delta t_\mathrm{c}$ for a 10-orbit binary for different scalar masses $\mu$ and initial densities $\rho_0$. Here $\bar{t}_\mathrm{c}$ is the merger (coalescence) time in the absence of a dark matter cloud. The effect is maximized for $\mu \approx 0.45 M^{-1}$, corresponding to a Compton wavelength of the dark matter particle that is comparable to the initial separation of the orbit $\lambda_c=2\pi/\mu\sim d_0\approx 12M$. 
    }
    \label{fig:tp_vs_mu}
\end{figure}

The DM energy densities required to give significant effects on the signal are high relative to the expected average galactic values, with the latter determined by large scale observations \cite{Pato:2015dua,Nesti:2013uwa,Li:2020qva,DeMartino:2018zkx,Ablimit:2020gxw}. Therefore the impact of such effects may be expected to be small \cite{Barausse:2014tra}. 
However, average galactic densities describe DM on large scales only, and its distribution on small scales (in particular the parsec and below scales relevant for astrophysical BHs) is not well constrained \cite{Benito:2020lgu}. There exist several mechanisms that could create DM overdensities around isolated BH. One well known possibility is the superradiant instability, in which a bosonic field extracts energy and angular momentum from a highly spinning black hole via repeated scattering in the ergoregion  \cite{1971JETPL..14..180Z,Press:1972zz,Zouros:1979iw,Detweiler:1980uk,Cardoso:2004nk,East:2017ovw} (see \cite{Brito:2015oca} for a review).
Another more prosaic effect is simply the accretion of dark matter in the potential well around BHs, which results in the formation of  ``dark matter spikes'' \cite{Gondolo:1999ef}
(a combination of both superradiance and accretion may lead to even higher densities \cite{Hui:2022sri}). 
Such spikes were originally proposed in the context of WIMP-like dark matter, but in general their profile is a power law with an exponent that depends on the effective equation of state of the dark matter \cite{DeLuca:2023laa,Berezhiani:2023vlo,Sadeghian:2013laa,Gnedin:2003rj,Merritt:2006mt,Merritt:2003qk,Shapiro:2022prq}. However, they also occur for low mass, wave-like DM candidates, with a form that is dependent on the relative Compton wavelength of the DM particle and the black hole horizon \cite{Clough:2019jpm,Hui:2019aqm,Bamber:2020bpu,Bucciotti:2023bvw,deCesare:2023rmg,Sanchis-Gual:2016jst}. In both cases, the DM density near the BHs depends on the asymptotic dark matter environment and on the particle properties.

However, a key question is whether these overdensities around isolated objects persist during a binary merger event. In the case of heavy (particle-like) DM \cite{Bertone:2004pz}, N-body simulations have shown that they disperse for equal mass mergers, meaning that objects close to merger or with a violent merger history are likely to have lost their DM environment \cite{Merritt:2002jz,Bertone:2005hw,Kavanagh:2018ggo}. Dark matter spikes nevertheless remain relevant for intermediate and extreme mass ratio inspirals (IMRIs and EMRIs) or primordial black holes, with signatures potentially detectable in next generation space and ground based observations \cite{Ferreira:2017pth,Eda:2013gg,Coogan:2021uqv,Cole:2022ucw,Kavanagh:2020cfn,Hannuksela:2019vip,Polcar:2022bwv,Amaro-Seoane:2007osp,Jangra:2023mqp,Li:2021pxf,Yue:2019ozq,Yue:2017iwc,Becker:2021ivq,DeLuca:2021ite,DeLuca:2022xlz,Takahashi:2023flk,Barsanti:2022vvl,Kim:2022mdj}.
For light or wave-like DM \cite{Schive:2014dra} (see \cite{Hui:2021tkt,Urena-Lopez:2019kud,Niemeyer:2019aqm} for reviews), much work has focused on the impact of black holes moving in galactic DM halos \cite{2023arXiv231103412K,Boudon:2023qbu,Zhong:2023xll,Cardoso:2022nzc,Cardoso:2022vpj,Vicente:2022ivh,Annulli:2020lyc,Annulli:2020ilw,Brax:2019npi} or with superradiant clouds \cite{Cao:2023fyv,Traykova:2023qyv,Chia:2022udn,Baumann:2018vus,Baumann:2019ztm,Ikeda:2020xvt,Cardoso:2020hca,Hannuksela:2018izj,Baumann:2022pkl,Zhang:2019eid,KumarPoddar:2019jxe,Wong:2019kru,Wong:2020qom,Kavic:2019cgk,Chia:2023tle,East:2018glu,Siemonsen:2019ebd,Tsukada:2020lgt,Baumann:2021fkf,Siemonsen:2022yyf,Tomaselli:2023ysb}. Some of this work has suggested that the cloud is not completely lost. In a previous publication \cite{Bamber:2022pbs}, we demonstrated that overdensities around equal mass binaries grew into a quasi-stationary profile that persisted up until the merger (see also \cite{Ficarra:2021qeh,Zhang:2022rex,Choudhary:2020pxy,Yang:2017lpm}).

In this Letter, we build on our study to better understand how generic such an effect is, and to properly quantify the impact that the DM has on the binary close to merger. We focus on the effect of wave DM on equal mass BH mergers, and in particular its dependence on the mass of the scalar particle. We simulate a 10-orbit binary black hole in an initially homogeneous dark matter environment starting from initial conditions satisfying the Hamiltonian and momentum constraints. We identify the decay of the orbit (and, as a consequence, dephasing of the gravitational wave signal) as being a direct result of the scalar cloud.  Our key results are illustrated in Fig. \ref{fig:tp_vs_mu}, where we show the dephasing is maximized when the mass of the scalar particle is such that its Compton wavelength is comparable to the initial separation of the orbit $\lambda_c=2\pi/\mu \sim d_0$. In addition, we are able to quantify the different channels that contribute to the dephasing in our scenario, finding the dominant effect to be driven not by radiation or dynamical friction drag forces, as are often discussed, but rather the attraction of the binary to the central overdensity.\\


\label{sec-background}
\textbf{\textit{Key background and physical setup.}}— We consider a minimally coupled massive complex scalar field $\Phi$ described by the action
\begin{equation}
    S = \int \dd^4 x \sqrt{-g}\left(\frac{R}{16\pi G} - \frac{1}{2}\left(\nabla_{\mu}\Phi\right)^* \left(\nabla^{\mu}\Phi\right) - V(\Phi)\right),
\end{equation}
with a quadratic potential $V(\Phi) = \mu^2 \Phi^* \Phi/2$, where $\mu$ is the scalar field mass. The dynamics of the scalar field is given by the Klein-Gordon equation coupled to gravity
\begin{equation}
\left[\nabla^{\alpha}\nabla_{\alpha} - \mu^2\right]\Phi = 0\,.
\end{equation}

\begin{figure}[t]
    \centering
    \href{https://youtu.be/2VJIfqCp7D8}{
    \includegraphics[width=\linewidth]{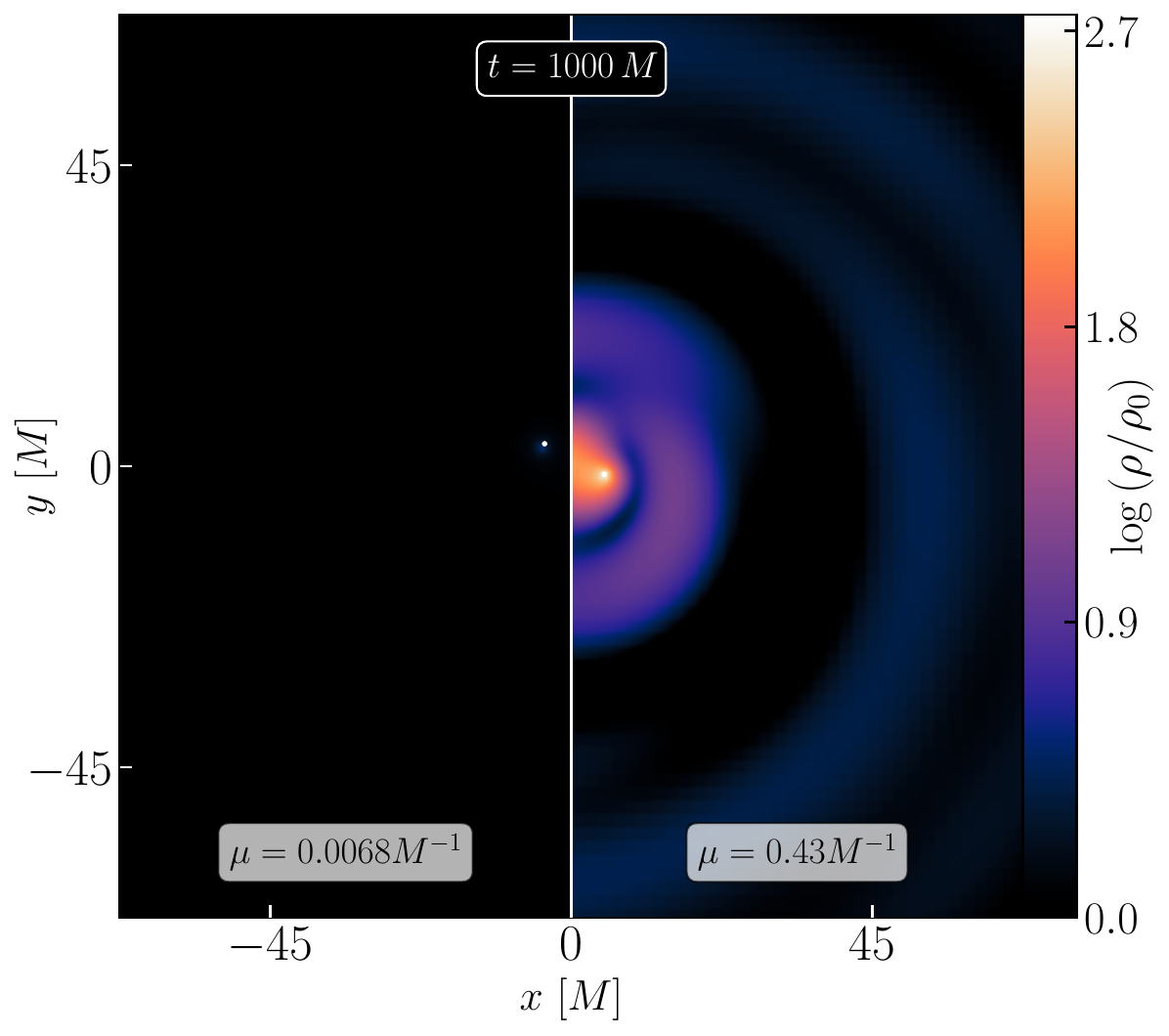}}
    \caption{Cloud density for two values of $\mu$: one large with $\lambda_c\approx d_0$ (right) in which case the binary obtains an enhanced density cloud, and one smaller, $\lambda_c\gg d_0$ (left) in which case the pressure coming from the long wavelength of the collective excitations of the field prevents a high density cloud from forming. We refer to these as the ``cloud'' and ``no cloud'' cases, respectively. Simulation movie \cite{movie}.}
    \label{fig:binary}
\end{figure}

In the case of a single BH immersed in a reservoir of such scalar DM, the stationary solution near the black hole is described by the Heun functions \cite{Hui:2019aqm,Santos:2020sut,Vieira:2014waa,Hortacsu:2011rr}, with a power law envelope and characteristic oscillations in the spatial profile on length scales set by the scalar wavelength. In the case of a binary no analytic form for a stationary state is known, but simulations \cite{Bamber:2022pbs} using the numerical codes \textsc{grchombo} \cite{Andrade:2021rbd} and \textsc{grdzhadzha} \cite{Aurrekoetxea:2023fhl} have indicated that for a range of initial configurations and within a few orbits, the scalar matter evolves into a persistent quasi-stationary profile with density spikes near the black holes and an overdensity between them. 

Ideally we would set this ``natural'' quasi-stationary DM configuration as an initial condition, and study the impact the cloud has on the binary merger using general relativity. However, even if an analytic form was known, a consistent solution of the GR constraints would lead to changes to the initial effective masses and momenta of the black holes for different densities and profiles, making comparisons of the subsequent evolutions difficult to interpret. In particular, it is hard to know if the additional dephasing is arising due to matter effects or due to the increased initial eccentricity of the orbits. One can mitigate this by applying eccentricity reduction schemes to the initial data, but the fact that the clouds can be very dense near the horizon makes this challenging as the eccentricity is extremely sensitive to small changes.

In this Letter we take a simpler approach. Given the short relaxation timescale of the cloud ($\sim 2$ orbits), compared to the timescale of the merger we are simulating ($\sim 10$ orbits), we start all simulations from a homogeneous configuration with fixed initial density $\rho_0 = \mu^2\phi_0^2$ and allow the cloud to build up dynamically during the simulation. To do so, we choose homogeneous initial conditions for the real and imaginary components of the scalar field, $\Phi=(\phi_0,0)$ and $\partial_t\Phi=(0,\mu\phi_0)$.
As we vary $\mu$ (which gives us different cloud configurations, Fig. \ref{fig:binary}), we adjust $\phi_0$ so as to we keep the initial density $\rho_0$ unchanged, thus the initial trajectories of the binary as a result of solving the initial data will be the same in all cases, which allows comparison between different masses (and different scalar cloud profiles) in a more controlled way. See the Supplemental Material for more details about the numerical implementation of the evolution equations and initial data.\\

\begin{figure}[t!]
    \centering
    \href{https://youtu.be/2VJIfqCp7D8}{
    \includegraphics[width=\linewidth]{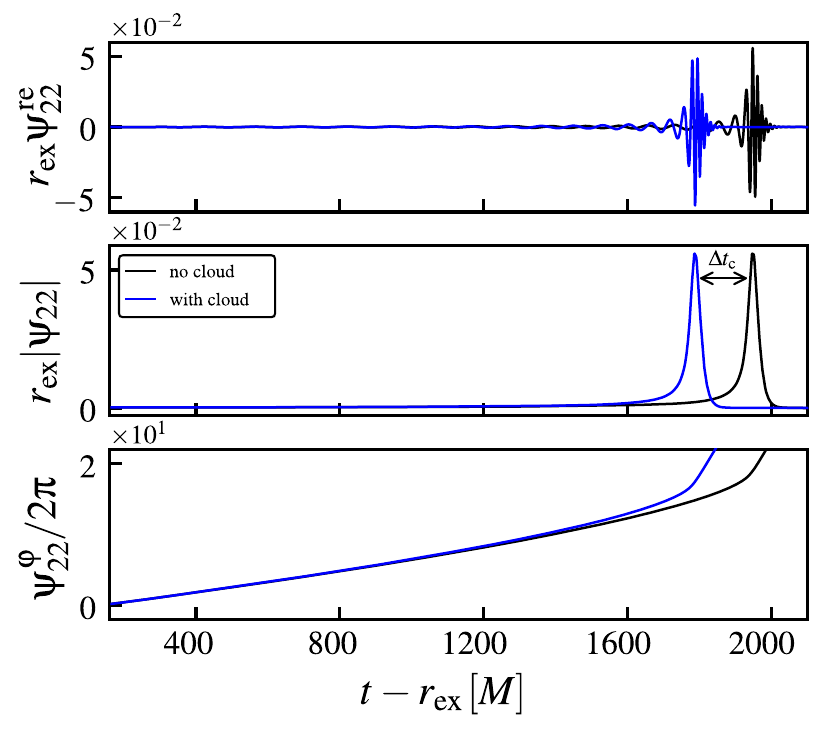}}
    \caption{Dephasing of the gravitational wave signal due to the accretion, dynamical friction and emission of wave dark matter around a binary black hole merger. Top panel is the real part of the $\psi_{22}$ mode, whilst mid and bottom panels are its modulus and phase, respectively. The black solid line corresponds to $(\mu,\phi_0)=(68,50)\times 10^{-4}$, which we refer to as ``no cloud'', as the Compton wavelength of the scalar field is much larger and we do not efficiently excite a DM cloud. The blue solid line corresponds to $(\mu,\phi_0)=(4300, 0.79)\times 10^{-4}$, and causes a $\Delta t_\mathrm{c}/\bar{t}_\mathrm{c} \approx 10\%$ dephasing of the merger time. The initial densities for both these cases is $\rho_0\approx 10^{-9} M^{-2}$. Movie in \cite{movie}.} 
    \label{fig:tensor_waves}
\end{figure}

\textbf{\textit{Dephasing of the binary.}}— 
We study the tensor gravitational-wave modes emitted by the binary black hole extracting the Newman-Penrose scalar $\Psi_4$ with tetrads proposed by \cite{Baker:2001sf}, projected into spin-weight $-2$ spherical harmonics,  $\psi_{lm}=\oint_{S^2}\Psi_4|_{r=r_\mathrm{ex}}\left[{}_{-2}\bar{Y}^{lm}\right]\,\rmd \Omega$, 
where $\rmd \Omega = \sin\theta\,\rmd\theta\,\rmd\varphi$ is the area element on the $S^2$ unit sphere.
The merger or coalescence time for our 10-orbit binary in vacuum is $\bar{t}_\mathrm{c}\approx 2000\,M$, defined as when $\vert\psi_{22}\vert$ peaks, which is the dominant mode. This is also the case for small initial densities $\rho_0$ since there is less backreaction of the matter on the binary metric. For a given density, a smaller effect is also seen for masses $\mu \ll M^{-1}$, as the Compton wavelength $\lambda_c\gg d$ and the cloud is not efficiently excited, see Fig. \ref{fig:binary}. We refer to the case of small $\mu$ as the ``no cloud'' configuration (it has the same initial, non zero density, but no structure forms around the binary), and the higher $\mu$ case as ``with cloud''. A typical result can be seen in Fig. \ref{fig:tensor_waves}. As expected, the presence of wave dark matter around the binary results in a dephasing of the gravitational-wave signal $\Delta t_\mathrm{c}\equiv t_\mathrm{c}-\bar{t}_\mathrm{c}$, which is caused by effects like accretion and dynamical friction from the cloud. We will give explanations and order of magnitude estimates for the various effects in the following section.

In Fig. \ref{fig:tp_vs_mu} we compare the dephasing for different DM masses $\mu \in \{0.0068,\, 0.86\}M^{-1}$, corresponding to wavelengths $\lambda_c \in \{924,\, 7\}M$ that span a range above and below the initial binary separation $d_0\approx 12 \,M$. We find that the dephasing is maximized for $\mu \approx 0.45 M^{-1}$, corresponding to $\lambda_c\approx 14 M \approx d_0$.  If the mass is smaller $\mu < 0.45 M^{-1}$, the cloud quickly becomes suppressed and the dephasing becomes negligible, but we note that larger separations earlier in the lifetime of the binary may support clouds at smaller masses. If the mass is larger $\mu > 0.45 M^{-1}$, the dephasing is smaller but remains significant, and we still find an efficient excitation of the cloud. On the one hand, this is not so surprising -- even at our highest mass, we are still in a regime where $\mu \approx M^{-1}$, and so as the merger radiates gravitational waves and inspirals in, it eventually approaches an orbital separation comparable to $\lambda_c$.
On the other hand, in other studies (see e.g. \cite{Traykova:2023qyv}) one often finds that the behaviour at this limit is already reasonably well described by the particle limit, and so we might have expected to see a greater dissipation of the cloud and suppression of the effect. The fact that this is not the case implies that the mass does not need to be very finely tuned for the effects to be significant, and motivates a more detailed study to find the boundary between the wave and particle regimes.

We also vary the asymptotic energy density to find the value at which the dephasing is detectable in our simulations during the last 10 orbits, which gives an indication of the value required for effects to be significant at merger.  See the conclusion section for these values in physical units.\\ 

\textbf{\textit{Quantification of the causes of the dephasing.}}—
To quantify the origins of the dephasing we identify the changes in the energy, angular and radial momentum of the binary that relate to the presence of the cloud of matter. 
We follow the approach of \cite{Clough:2021qlv,Croft:2022gks,Traykova:2021dua,Traykova:2023qyv}, and define a current $J^\mu=\xi^\nu T^\mu_\nu$ in the direction $\xi^\nu$ and associated charge and flux
\begin{equation}
    Q =-n_\mu J^\mu \qquad F= \alpha N_i J^i\,,
\end{equation}
where $N_i$ is the outward normal direction to the surface that bounds the volume. If $\xi^\nu$ is a Killing vector then $\nabla_\mu J^\mu=0$ and the change in charge is balanced by a flux through a surface. When this is not the case the conservation laws require an additional ``source'' term
\begin{equation}
    S = \alpha T^\mu_\nu \nabla_\mu \xi^\nu\,,
\end{equation}
describing the exchange of the charge between matter and curvature. It is this quantity that corresponds to gravitational forces in the Newtonian limit, and that quantifies the way in which momentum is extracted from the binary by the matter\footnote{We also include in this quantity the flux of the matter charge into spheres around the BHs, i.e. the accretion of the matter energy and momenta into the BHs, as in \cite{Traykova:2023qyv, Traykova:2021dua}.}. The explicit expressions for the energy, angular and radial momentum charges, sources and fluxes in terms of the ADM variables are given in the Supplemental Material. In Fig. \ref{fig:fluxes} we plot the time integration of each of these quantities and verify their agreement. 
In each case the black line should equal the sum of the blue and red lines and provides a check on the error. It is the red line that quantifies the exchange of the relevant charge from the binary to the matter, and therefore that drives the dephasing. 

\begin{figure}[t]
    \centering
    \includegraphics[width=\linewidth]{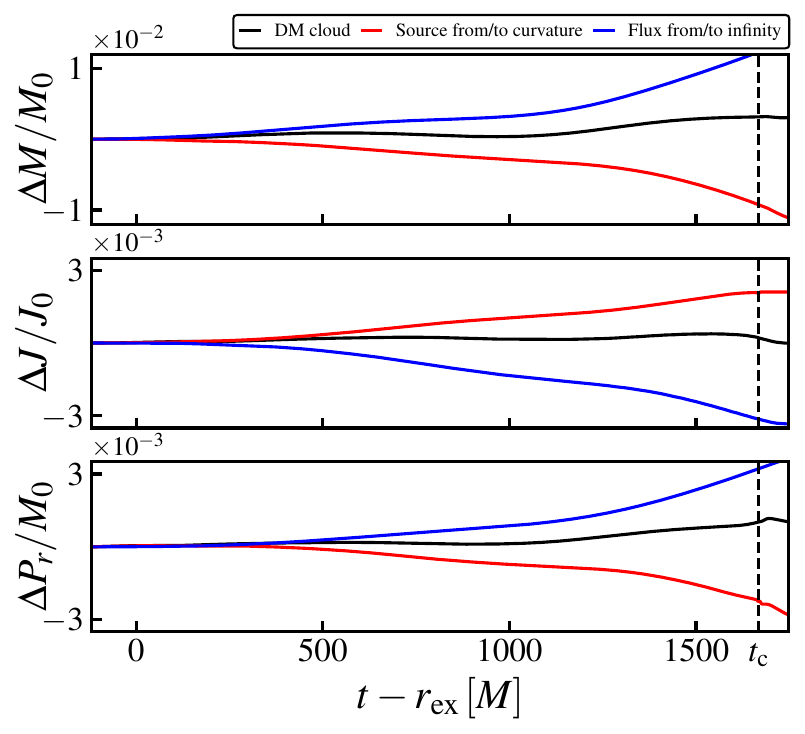}
    \caption{Conservation law for the energy, angular (corotating) and radial (inward) momentum of the matter in a sphere of radius $40M$ around the binary that contains the main cloud overdensity (Fig. \ref{fig:binary}). The black line shows the change in each respective charge for the cloud over time and is the sum of the other two lines.
    The red line describes the time integration of the exchange of the charge from the curvature to the matter.The blue line is (minus) the total flux of the charge into the outer bounding surface.}
    \label{fig:fluxes}
\end{figure}

In the top panel of Fig \ref{fig:fluxes}, we see that the {\it energy} of the matter within the volume increases, due to the flux of matter energy across the outer surface -- this is simply reflecting the fact that the central cloud density grows over time due to accretion from the environment. The increase is partially offset by a negative source term, which is mainly driven by the accretion of energy into the BHs, increasing their masses by approximately 1\% over the course of the merger (we can check that this agrees to the change in their measured masses from the apparent horizon finder). After the merger, the energy in the cloud around the remnant decreases slightly as some is accreted onto the remnant, but it does not completely dissipate.

In the second panel, the {\it angular momentum} held in the matter cloud initially increases as the curvature of the binary ``stirs up" the cloud during the transient phase, then reaches a reasonably steady state during which the rate of extraction of angular momentum from the spacetime curvature (the stirring) is balanced by its flux out of the outer surface. The result is that the angular momentum of the binary is decreased, and carried away by scalar waves. We can view the source from/to curvature as a dynamical friction effect -- the extraction of the angular momentum of the spacetime of the binary by the matter.  After merger there is an increased flux of radiation from the outer surface of the volume which carries away all the angular momentum built up in the cloud during the inspiral.

In the third panel, the {\it radial momentum} held in the matter cloud is tracked. Here, we see that the matter cloud overall gains some inward radial momentum. However, the accretion of inward radial momentum from the outer surface is roughly balanced by its loss into curvature -- partly as a result of the binary accreting the ingoing momentum, and partly as a result of it being attracted to the central overdensity. This gives the binary an inward pull that accelerates during the final plunge.

These measurements allow us to quantify which of the three contributions accounts for most of the dephasing. For circular orbits the angular momentum is given by $J=M r^2\left(2\pi/T\right)$ and very roughly we can say that the total dephasing $\Delta t_c /\bar{t}_c \sim n\Delta T_0/(nT_0)$, where $n$ is the number of orbits and $T_0$ is the period of the first orbit. The changes in $T$ can be estimated in terms of the total changes of mass, angular momentum and radius (expressed in terms of the final amplitudes of the red lines in Fig. \ref{fig:fluxes}) as
\begin{align}
    \frac{\Delta t_c}{\bar{t}_c} \approx \frac{n\Delta T}{n T_0} \approx \frac{\Delta M^\mathrm{tot}}{n M_0} - \frac{\Delta J^\mathrm{tot}}{n J_0} + \frac{2\Delta P_r^\mathrm{tot}\, \bar{t}_c}{n M_0 r_0 }
\end{align}
where we have used $\Delta r/r_0 \approx \Delta P_r^\mathrm{tot}\, \bar{t}_c^2 /(r_0 M_0 \bar{t}_c)$ in the last term, obtained from integrating the inward radial force on each BH over the inspiral time $\bar{t}_c$. Using that $r_0=d_0/2$, $n\approx 10$, $\bar{t}_c\approx 2000M$ and the values observed in Fig. \ref{fig:fluxes}, we estimate $\mathcal{O}(0.1)\%$, $\mathcal{O}(0.01)\%$ and $\mathcal{O}(10)\%$ for each term, respectively. We conclude that the radial force towards the overdensity is the dominant cause of the dephasing.\\


\textbf{\textit{Conclusion.}}— Using general relativistic simulations of a binary accreting dark matter, we have shown that the dephasing in the gravitational-wave signal of an equal mass black hole merger is maximized when the Compton wavelength of the dark matter particle is comparable to the orbital distance of the binary, $2\pi/\mu\sim d$. 
We need the mass of the scalar to be sufficiently large for a central overdensity to build up -- low mass scalars suppress structure on smaller scales than their Compton wavelength. Converting into physical units, the optimal scalar mass to induce dephasing in the last 10 orbits of an equal mass binary with total mass $M$ is then $
   \mu \approx 5\times 10^{-17}\, \left(M/10^6 M_\odot\right)^{-1}  \,\mathrm{eV}$,
which can result in a $10\%$ dephasing during the last 10 orbits of the binary (taking the blue line in Fig. \ref{fig:tp_vs_mu}) for asymptotic densities around the BH of
\begin{equation}
    \rho_0 \approx 10^{20} \,\left(\frac{M}{10^6 M_\odot}\right)^{-2}\,\frac{M_\odot}{\mathrm{pc}^3}\,.
\end{equation}
This is high relative to the average DM density, but our measured dephasing is only over a short period of the binary's lifetime ($\sim 10$ orbits), and has a cumulative effect. Therefore smaller densities could give sufficient dephasing to be detectable assuming the effect is triggered at larger separations (which would also allow lower mass candidates to contribute to the effect), or if observations can happen over a longer time frame (e.g. by combined LISA/LVK observations).

The simulations in this work demonstrate that accumulation of wave-like dark matter between the binary could have a significant effect on the merger history of binaries, unlike in particle cases where the dark matter tends to disperse. As recently suggested in \cite{2023arXiv231103412K}, it could even go some way to explaining the final parsec problem. In particular, we highlight the importance of considering the radial force arising from any central overdensity that forms, in addition to the radiation of waves carrying angular momentum and energy away from the binary. 
As noted above, the effects remain significant even at the higher end of the masses that we can probe in our simulations, at which $\mu M \approx 1$. Further investigations should be made to determine the point at which particle-like behaviour takes effect, and to study the importance of the relativistic features in our simulations such as the presence of black hole horizons.\\


\textbf{\textit{Acknowledgements.}}— We would like to thank Jean Alexandre, Emanuele Berti, Gianfranco Bertone, Robin Croft, Giuseppe Ficarra, Thomas Helfer, Charlie Hoy, Lam Hui, Macarena Lagos, Eugene Lim, Miren Radia, Dina Traykova, Rodrigo Vicente, Sebastian von Hausegger and Helvi Witek for helpful conversations. We thank the GRChombo collaboration (\href{www.grchombo.org}{www.grchombo.org}) for their support and code development work. JCA acknowledges funding from the Beecroft Trust and The Queen's College via an extraordinary Junior Research Fellowship (eJRF). KC acknowledges funding from the UKRI Ernest Rutherford Fellowship (grant number ST/V003240/1). JB acknowledges funding from a Science and Technology Facilities Council (STFC) PhD studentship and funding from National Science Foundation
(NSF) Grant PHY-2006066. PGF acknowledges support from STFC and the Beecroft Trust.

This work was performed using the DiRAC@Durham
facility managed by the Institute for Computational Cosmology on behalf of the STFC DiRAC HPC Facility (www.dirac.ac.uk) under DiRAC RAC15 Grant ACTP316. The equipment was funded by BEIS capital funding via STFC capital grants ST/P002293/1, ST/R002371/1 and ST/S002502/1, Durham University and STFC operations
grant ST/R000832/1.
Part of this work was performed using the DiRAC Data Intensive service at Leicester, operated by the University of Leicester IT Services, which forms part of the STFC DiRAC HPC Facility (www.dirac.ac.uk). The equipment was funded by BEIS capital funding via STFC capital grants ST/K000373/1 and ST/R002363/1 and STFC DiRAC Operations Grant ST/R001014/1. This work also used the DiRAC@Durham facility managed by the Institute for Computational Cosmology on behalf of the STFC DiRAC HPC Facility (\href{www.dirac.ac.uk}{www.dirac.ac.uk}). The equipment was funded by BEIS capital funding via STFC Capital Grants ST/P002293/1, ST/R002371/1 and ST/S002502/1, Durham University and STFC Operations Grant ST/R000832/1. DiRAC is part of the National e-Infrastructure.

\bibliography{mybib}

\newpage
\clearpage
\appendix

\section*{Numerical implementation, diagnostics and convergence tests}

We evolve the gravity sector solving the Einstein field equations for a line element that we decompose in the usual ADM form \cite{Arnowitt:1962hi}
\begin{equation}
    ds^2 = -\alpha^2 \rmd t^2 + \gamma_{ij}(\rmd x^i + \beta^i \rmd t)(\rmd x^j + \beta^j \rmd t),
\end{equation}
where $\gamma_{ij}$ is the three-dimensional spatial metric that we decompose into a conformal factor and a conformally related metric $\gamma_{ij}=\bar{\gamma}_{ij}/\chi$. The lapse and shift gauge functions $\alpha$ and $\beta^i$ determine the choice of spatial hyperslicings and their coordinates, which in numerical relativity are dynamically determined. The extrinsic curvature tensor $K_{ij}=(2 D_{(i} \beta_{j)}-\partial_{t} \gamma_{ij})/2\alpha$ is decomposed into a trace $K$ and a traceless part $A_{ij}$, i.e. $K_{ij} = A_{ij}+(1/3)K\gamma_{ij}$. We evolve use the CCZ4 formulation \cite{Alic:2011gg} and the moving puncture gauge \cite{Bona:1994dr,Baker:2005vv,Campanelli:2005dd,vanMeter:2006vi} with \textsc{grchombo} \cite{Andrade:2021rbd,Radia:2021smk,Clough:2015sqa}.

\begin{figure}[t!]
    \centering
    \includegraphics[width=\linewidth]{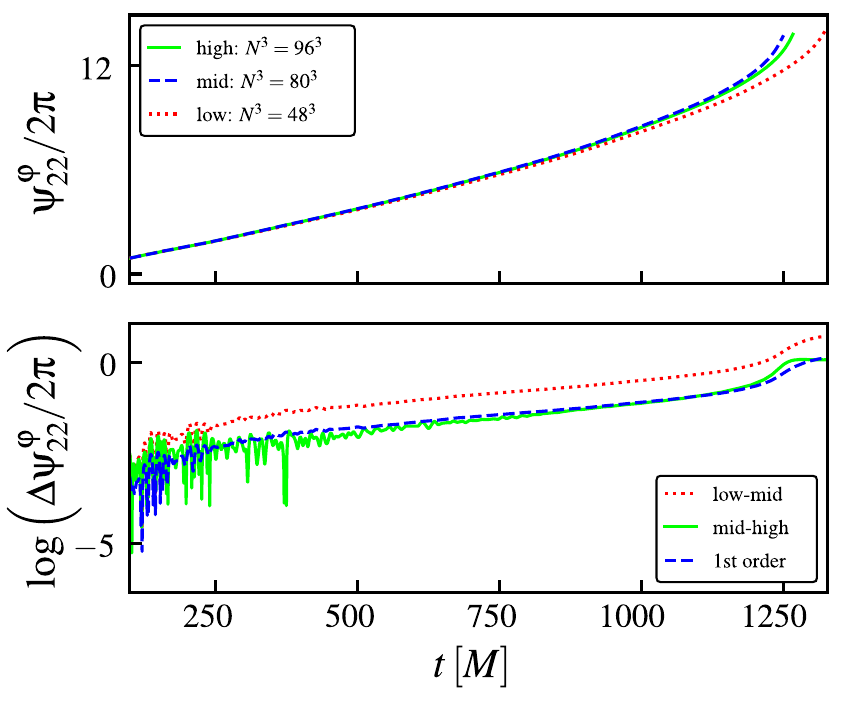}
    \caption{Convergence testing of the gravitational-wave phase evolution for the largest dephasing case: $(\mu,\phi_0)=(4300, 2.5)\times 10^{-3}$, so that the initial density is $\rho_0 \approx 10^{-8} M^{-2}$. We use three different resolutions with $N^3$ number of grid points on the coarsest level. The decrease in the error is consistent with $1^\mathrm{st}$ order.}
    \label{fig:Relativistic_conv_test}
\end{figure}

We solve the Hamiltonian constraint using Bowen-York initial data using the CTTK hybrid method \cite{Aurrekoetxea:2022mpw} (see table \ref{table:BH_params} for the binary parameters). In the homogeneous limit, this reduces to choosing the trace of the extrinsic curvature tensor $K^2=24\pi G \rho$ and solving for a correction of the conformal factor $\chi$ sourced by the traceless components $A_{ij}$.
We choose $K<0$ so that the scalar field is initially decaying, which chooses the more conservative impact on the merger. The value is small and the effect of either choice on the overall trends observed is not significant.
We use a simulation box length $L=512M$ and $9$ levels of mesh refinement (See Fig. \ref{fig:Relativistic_conv_test} for convergence tests). We impose reflecting boundary conditions at $z=0$, while for the other boundaries we impose either zeroth order extrapolating boundary conditions (matching the values on the exterior ghost cells to the value (radially directed) inside the simulation grid) or Sommerfeld boundary conditions.

\begin{table}[b!]
\begin{center}
\setlength{\tabcolsep}{0.7em}
\begin{tabular}{|c||c|c||c|} 
     \hline \hline
     $d/M$  &   $12.21358$ & $\vert p_x\vert/M$   &     $5.10846\times 10^{-4}$ \\
     $\mbh/M$   &   $0.48847892320123$  & $\vert p_y\vert/M$   &     $8.41746 \times 10^{-2}$\\
     $T/M$     &      $271.34$ & $\vert p_z\vert/M$   &      $0$\\
     \hline \hline
\end{tabular}
\end{center}
\caption{Black hole binary initial parameters. The black holes are initially aligned along the $x$ axis in the $z=0$ plane, with initial momenta $\vec{p}_1 = (-\vert p_x\vert,+\vert p_y\vert,0)$ for the BH with initial position $\vec{r}_1=(d/2,0,0)$ and $\vec{p}_2 = (+\vert p_x\vert ,-\vert p_y\vert,0)$ for the one at $\vec{r}_2=(-d/2,0,0)$.}
\label{table:BH_params}
\end{table}

Following the approach of \cite{Clough:2021qlv,Croft:2022gks}, the energy momentum tensor is decomposed into the energy density, momentum density and stress-energy density measured by normal observers as
\begin{equation}
    T^{\mu\nu} = \rho n^\mu n^\nu +  S^\mu n^\nu + S^\nu n^\mu + S^{\mu\nu}.
\end{equation}
The conservation law is written as 
\begin{equation}
    \partial_t \left(\int Q \mathrm{d}V\right)= \int S \mathrm{d}V -  \int F \mathrm{d}A\,,
\end{equation}
where these correspond to the change in charge in the cloud, the exchange between matter from/to curvature, and the flux to/from infinity. The energy and angular momentum currents are related to the time-like, angular and radial directions $J^\mu_t = T^\mu_\nu \xi^\nu_t$, $J^\mu_\phi = T^\mu_\nu \xi^\nu_\phi$ and $J^\mu_r = T^\mu_\nu \xi^\nu_r$, where $\xi_t^\nu = (1,0,0,0)$, $\xi_\phi^\nu=(0,-y,x,0)$ and $\xi_r^\nu=-(0,x,y,z)/r$. The respective charges $Q$ and fluxes $F$ in terms of ADM quantities are then 
\begin{align}
    Q_t &= -\alpha \rho +\beta_k S^k\\
    F_t &=  N_i\left(\beta^i(\alpha \rho - \beta^j S_j) + \alpha(\beta^k S_k^i - \alpha S^i\right)\\
    Q_{\lbrace{\phi,r\rbrace}} &= S_i \xi^i_{\lbrace{\phi,r\rbrace}} \\
    F_{\lbrace{\phi,r\rbrace}} &= -N_i \beta^i S_j \xi^j_{\lbrace{\phi,r\rbrace}} + \alpha N_i S^i_j \xi^j_{\lbrace{\phi,r\rbrace}} \,,
\end{align}
where $N_i=(x,y,z)/r$ is the normalised radial unit vector, with $s_i=(x,y,z)/r$ and $N_i = s_i / \sqrt{(\gamma^{jk} s_j s_k)}$.
The source terms

\begin{align}
    S_t =& -\rho \partial_t \alpha + S_i \partial_t \beta^i + \frac{\alpha}{2} S^{ij}\partial_t\gamma_{ij} \\
    S_{\lbrace{\phi,r\rbrace}} =& \alpha S^\mu_\nu \partial_\mu \xi^\nu_{\lbrace{\phi,r\rbrace}} + \alpha S^\mu_\nu {}^{(3)}\Gamma^\nu_{\mu\sigma} \xi^\sigma_{\lbrace{\phi,r\rbrace}} \nonumber \\ 
    & - S_\nu \beta^i \partial_i\xi^\nu_{\lbrace{\phi,r\rbrace}} + S_\nu \xi^\mu_{\lbrace{\phi,r\rbrace}} \partial_\mu\beta^\nu - \rho\xi^\mu_{\lbrace{\phi,r\rbrace}} \partial_\mu \alpha\,
\end{align}
where $\partial_t \gamma_{ij}=-2\alpha K_{ij} +D_i\beta_j + D_j \beta_i$, and both $\partial_t\alpha$ and $\partial_t\beta^i$ are given by our moving puncture gauge conditions. The quantities $\partial_\mu\xi^\nu$ are all zero except $\partial_x \xi^y = 1$ and $\partial_y \xi^x = -1$

We also track the flux through inner surfaces that move together with the black holes, which introduces additional advection terms to the flux
\begin{equation}
    F^\mathrm{BH} = \alpha N_i^\mathrm{BH}J^i - N_i^\mathrm{BH} \beta^i \left( Q -  S/2\right)\,,
\end{equation}
where $N_i^\mathrm{BH}$ is defined above.
\clearpage

\end{document}